\documentclass[a4paper,UKenglish]{lipics}
 
\usepackage{microtype}
\usepackage{cmdtrack}

\usepackage{multirow}
\usepackage{enumerate}

\newcommand{\init}{\varepsilon}
\newcommand{\On}{\mathrm{On}}

\newcommand{\calK}{\mathcal{K}}

\newcommand{\calC}{\mathcal{C}}
\newcommand{\calG}{\mathcal{G}}

\newcommand{\autW}{\mathcal{W}}

\newcommand{\xrarrow}[1]{\overset{#1}{\longrightarrow}}

\newcommand{\takeout}[1]{}

\theoremstyle{definition}

\title{A Retraction Theorem for Distributed Synthesis}
\author[1]{Dietmar Berwanger}
\author[1,2]{Anup Basil Mathew}
\author[2]{R. Ramanujam}
\affil[1]{LSV, CNRS \& Universit\'e Paris-Saclay, France}
\affil[2]{Institute of Mathematical Sciences, Chennai, India}

\authorrunning{D. Berwanger, A.\,B. Mathew and R. Ramanujam} 

\Copyright{Dietmar Berwanger, Anup Basil Mathew and R. Ramanujam} 


\serieslogo{}
\volumeinfo
  {Billy Editor and Bill Editors}
  {2}
  {Conference title on which this volume is based on}
  {1}
  {1}
  {1}
\EventShortName{}
\DOI{10.4230/LIPIcs.xxx.yyy.p}

\begin{document}

\maketitle

\begin{abstract}
We present a general theorem for distributed synthesis problems in
coordination games with $\omega$-regular objectives of the form:
If there exists a winning strategy for the coalition, then there
exists an ``essential'' winning strategy, that is obtained by a
retraction of the given one.
In general, this does not lead to finite-state winning strategies, but
when the knowledge of agents remains bounded, we can solve the
synthesis problem.
Our study is carried out in a setting where objectives are expressed
in terms of events that may \emph{not} be observable.
This is natural in games of imperfect information, rather than the
common assumption that objectives are expressed in terms of events
that are observable to all agents.
We characterise decidable distributed synthesis problems in terms of
finiteness of knowledge states and finite congruence classes induced
by them.
\end{abstract}

\section{Introduction}
In the theory of system design and verification, the synthesis problem
is formulated as a win/lose game between the system and a hypothetical 
opponent labelled as environment, and the solution to the problem is
a winning strategy for the system. When these are games of infinite
duration, and the winning condition is a set of infinite regular plays,
a central theorem of the subject asserts that it can be effectively decided
whether winning strategies exist for the system; moreover, when winning
strategies exist, they can be realised using finite memory. A rich theory
of such games has been built in the last couple of decades 
(\cite{McNaughton66,Rabin72,PnueliRos89,Henzinger05,Thomas09}).

In the design of systems with multiple components that work concurrently,
a similar question can be formulated. When doing so, it matters whether
the components work {\em cooperatively} towards system goals against the
adversary, or could have potentially conflicting goals despite which we 
would like the system to achieve its goals against the environment. 
Moreover, a major raison-d'etre of such systems is {\em distributedness}:
each component has access to partial information about the global system.
This is needed for reliability and isolation of faults. Such considerations
lead to the formulation of the synthesis problem as a win/lose game of
{\sf imperfect} information, again between the system and an environment
(which may of course have global / perfect information). Moreover, we
wish to synthesize strategies for each component acting independently,
and hence we have a problem of distributed synthesis in imperfect
information games, even in the restricted case of coordinating agents,
with identical payoffs to all components. Since components act concurrently
and independently it matters whether they all act {\em asynchronously},
each with its own clock, or {\em synchronously}, according to the ticks
of a common global clock.

That distributed synthesis for imperfect information games is hard is 
well-known: the literature is replete with results showing undecidability 
of the winning strategy question 
(\cite{PnueliRos90},\cite{FinkbeinerSch05},\cite{Janin07},\cite{BerwangerKai10}). 
These results are in some sense generic, and have to do with `forking'
of information among players, which loosely amounts to uncertainty of
a player about another's information. Thus game states implicitly
carry information on `epistemic' states of players involving their
knowledge / ignorance of each other's states, and then about each other's
knowledge and so on; this grows unboundedly, and is a crucial source
of undecidablity.

Coping with such undecidability, researchers have tended to place
{\sf structural} constraints on the system: on the architecture of
interaction among players (\cite{MadhuThiagu02},\cite{MuschollWal14}), constraining patterns of
interaction to be statically determined 
(\cite{MohalikWal03},\cite{GastinLerZei04}), 
constraining the players to a hierarchy (\cite{KupfermanVar01},\cite{BerwangerMV15}), and so on. 
Semantic subclasses have been studied, e.g. using a homomorphic
characterisation (\cite{BKP11}), a recurrent certainty condition 
(\cite{BerwangerM14a}), and so on. These models have concurrent components, but
no explicit communication between them; \cite{RamSim10} identifies public vs
private communication as a potential source of (un)decidability.

A striking feature of all these studies is that where decidable
subclasses are obtained, they invariably result in the `epistemic
states' referred to above being bounded, if not in the system 
evolution as such, but in the construction of winning strategies.
Such an observation leads us to a theoretical consideration of
the question: is there a way to abstractly characterise the 
existence of winning strategies in terms of how these `epistemic
states' grow, when they are finite, bounded etc. The general
problem being undecidable, there cannot be a recursive characterisation
of this form; yet, we might discern a general pattern, specialising
to specific bounds for structurally or otherwise constrained applications.

This is the project taken up in this paper. When we have a finite-state
presentation of a game of imperfect information, with finite-state
objectives, if we are told that the coalition of players, has a
(distributed) winning strategy against an environment, can we identify
the core of the strategy in terms of the `epistemic states' encountered,
and its underlying dynamic structure ? We answer this by proving a
{\sf retraction theorem}: if a winning strategy exists, then a
`small' quotiented strategy exists as a witness. In specific cases,
the retraction can be strengthened, to obtain bounds on the size
of witnesses, leading to decidability. Interestingly, these bounds
assert memory that is bounded as a function of bounds on epistemic
states, which can we consider bounded memory in mutual information.

A natural question is: why bother ? (Alternatively phrased: so what ?)
We consider this theorem as rudimentary methodology for distributed
synthesis in imperfect information games: information sets pose a
uniformity condition on strategies but at the same time cause
cascading uncertainty. With finite state labelling, these mutual
information states may be tamed as our `lateral' retraction suggests;
temporal dynamics is addressed using {\em progress measures}, as for
instance studied in \cite{Klarlund94,KlarlundK95} and \cite{VoegeJ00,Jurdzinski00}. 
The interaction between these
`lateral' and `vertical' folding (in finite state situations) is
generic to these systems. We illustrate with two small examples,
one that of 1-player information games, and another of hierarchies
of players. However, the idea is that retractions of this kind can
be constructed for a wide class of systems.

The study is carried out in a model that is as general as possible.
This helps us generalize earlier results. For instance, a common 
assumption in distributed synthesis is that objectives are expressed 
in terms of events that are observable to all agents. Despite imperfect 
information, 
each agent can monitor the outcome of the interaction by updating an 
automaton that finally accepts if the objective is met. This is clearly
against the spirit of distributedness, and our model can dispense with
such assumptions. Thus the theorems include systems where objectives 
are expressed in terms of of events that may \emph{not} be observable.

\section{The Model}
We model a synchronous distributed system as a game between a 
\emph{team} $N = \{1, \dots, n \}$ of players and one passive agent 
called \textit{Nature}.  The game is played in an infinite sequence 
of stages.  Each player $i \in N$ has a finite set $A^i$ of \emph{actions} 
among which he can choose. We denote by $\displaystyle A := 
\times_{i \in N}A^i$ the set of action profiles --- by \emph{profile} we 
generally mean a tuple $x=(x^i)_{i\in N}$ of elements, one for each player 
in the team.    The choices of Nature are called \emph{directions} and range
over a finite set $D$.

In each stage, every player~$i$ chooses an action $a^i \in A^i$, then Nature 
chooses a direction $d$.  Together, these choices determine a \emph{move} 
$\alpha = (a, d)$; we denote the set of possible moves by $\Gamma = A \times D$.  
A \emph{play} is an infinite sequence $\pi \in \Gamma^\omega$ of moves, 
and a \emph{history} is a finite sequence $\tau \in \Gamma^*$ of moves.

We are interested in a {\em finite state} presentation of the game,
and hence the information revealed to player~$i$ during the play
is described by a Mealy machine~$M^i$ that reads histories and outputs 
sequences of \emph{observations} from a finite alphabet~$B^i$.
The machine $M^i=(Q^i, \delta^i, \beta^i, q^i_{\init})$ is specified
by a finite set $Q^i$ of states, a transition function 
$\delta^i: Q^i \times \Gamma  \to Q^i$, an output function
$\beta^i: Q^i \times \Gamma  \to B^i$, and an initial state
$q_{\epsilon}^i$.  For an input history 
$\tau = \alpha_0 \alpha_1 \dots \alpha_{\ell-1}$,  the \emph{run} $\rho( \tau)$ 
of the machine 
is the sequence $q_0, q_1, \dots, q_\ell$ with $q_0 = q_\init$ 
and $q_{k+1} = \delta^i( q_k, \alpha_{k} )$ for all $k \le \ell$, 
and its \emph{output} $\beta^i(\tau)$ is the sequence of observations 
$\beta^i( v_0, \alpha_0 ) \beta^i( v_1, \alpha_1 ) \dots \beta^i( v_{\ell -1}, \alpha_{\ell - 1}) $.
We say that two histories $\tau, \tau' \in \Gamma^*$ are indistinguishable 
for player~$i$, and write $\tau \sim^i \tau'$, if the runs of $M^i$ and 
$\tau$ and $\tau'$ yield the same observation sequence 
$\beta^i( \tau ) = \beta^i( \tau')$.  Clearly, this is a synchronous 
equivalence relation.

Overall, a distributed game is specified by a structure 
$\calG = ( \Gamma^*, (\sim^i)_{i \in N} )$ on the tree of histories ordered by the word prefix relation  
with a profile of indistinguishability relations described by finite-state Mealy 
machines.

A \emph{decision structure} for the game $\calG$ is a directed graph 
$S = (V, E, f, v_0)$ on a possibly infinite set of nodes $V$ with an edge 
relation $E \subseteq V \times D \times V$ that describes a deterministic 
transition function from $V \times D$ to $V$, an action choice function 
$f: V \to A$, and an initial node $v_\init$.  
Every path $v_0 \xrarrow{d_0} v_1 \xrarrow{d_1} \dots
\xrarrow{d_{\ell-1} }v_\ell$ in $S$ starting from $v_0 = v_\init$ 
identifies a unique history $f( v_0 ) d_0 f(v_1) d_1 \dots f(v_{\ell -1}) d_{\ell - 1}$ 
in~$\calG$. We say that a history~$\tau$ \emph{follows} the decision structure~$S$, 
or simply $\tau$ is in~$S$, if $\tau$ is identified by a path in $S$. 
Conversely, every history 
$\tau = a_0 d_0 \, a_1 d_1\, \dots\, a_{\ell-1} d_{\ell-1}$ in $\calG$ 
corresponds in~$S$ to a unique path 
$v_0 \xrarrow{d_0} v_1 \xrarrow{d_1} \dots
\xrarrow{d_{\ell-1} }v_\ell$ from the
initial node $v_0 = v_\init$.
We call the sequence $v_0 v_1 \dots v_\ell$ the \emph{trace} of~$\tau$.
Notice that $\tau$ follows~$S$ if, and only if,   
$a_k = f( v_k )$ for all $k \le \ell$;
in this case, we write $v_\init \xrarrow{\tau} v$ for the end node $v = v_\ell$.
For each player~$i$, 
the indistinguishability relation $\sim^i$ on~$\calG$ induces a \emph{uniformity} relation
among states of~$S$: we write $v \approx^i_{S} v'$ if there 
exist histories $\tau \sim^i \tau'$ such that 
$v_\init \xrarrow{\tau} v$ and $v_\init \xrarrow{\tau'} v'$. 

A \emph{strategy structure} is a decision structure that satisfies the following uniformity condition:
if $v \approx_{S}^i v'$ then $f^i( v ) = f^i( v' )$, for all $v, v' \in V$. 
The uniformity relation arises as an operational consequence of 
epistemic indistinguishability: if two nodes are reachable by indistinguishable histories, 
they prescribe the same action for all pairs of histories that reach them.  
Notice that $\approx_{S}^i$ 
is reflexive and symmetric, but not necessarily transitive.

Alternatively, we can view a strategy structure as a Moore machine 
(possibly on infinitely many states) that
implements a function $S: \Gamma^* \to A$ from histories to action profiles, 
by assigning to each history $\tau$, the action profile $f( v )$ prescribed at the state~$v$ reached 
by the path corresponding to $\tau$ in the structure~$S$. 
This assignment is
\emph{information-consistent} for each player~$i$: for any pair
$\tau \sim^i \tau'$ of 
indistinguishable histories that follow $S$,
the action profiles $S( \tau )$ and $S( \tau' )$ agree on
their player-$i$ component.
Accordingly, $S$ can also be implemented by a
distributed profile of private Moore machines $s^i$ each of which
inputs observation sequences $\tau^i := \beta^i( \tau )$ of player~$i$ 
and return actions $s^i( \tau^i )$ such that 
$S( \tau ) = (\, s^i( \beta^i(\tau))\, )_{i \in N}$ for all histories $\tau$ in $S$.

We are interested in strategies that enforce a specified branching-time 
behaviour of distributed systems.  Accordingly, we define the 
\emph{outcome} of a strategy $S$ in a game~$\calG$ to be the tree
$H_S \subseteq \Gamma^*$ in $S$ equipped with 
the prefix order.  To specify $\omega$-regular properties 
of strategy outcomes, we use tree automata $\autW:=(Q,\Delta, q_{\epsilon},\Omega)$
where $Q$ is a finite set of states, $\Delta \subseteq Q \times A \times Q^D$ 
is a nondeterministic transition function and $\Omega:Q \to \mathbb{N}$ is a
labelling of the states with \emph{priorities} from a finite range 
$\Omega( Q ) \subseteq \mathbb{N}$. 
A \textit{run} of the automaton $\autW$ on (the outcome of) a 
strategy $S$ is a labelling $\rho: H_S \to Q$ of the outcome tree
such that $\rho(\epsilon)=q^\epsilon$ and, for all histories $\tau \in
H_S$ with $a := S( \tau)$, we have
  $(\rho(\tau), a, (\rho( \tau ad ))_{d \in D} ) \in \Delta$.
The run $\rho$ is \textit{accepting} if, for every play
$\alpha_0 \alpha_1 \dots$ through $H_S$, the corresponding priority sequence
$\Omega( \rho(\epsilon)),\Omega( \rho(\alpha_0)), \Omega( \rho(\alpha_0 \alpha_1)), \dots$ satisfies
the \emph{parity condition}, which requests that the
minimum priority appearing infinitely often in the sequence be even. 
We say that a joint strategy $s$ is \emph{winning} for $\autW$, if
there exists an accepting run of $\autW$ on the outcome tree of $S$. 
 
We consider the following \emph{distributed synthesis problem}: 
Given a distributed game $\calG$ and a specification $\autW$, 
decide whether a finite-state winning strategy exists, and if so, 
construct one.
Following \cite{PetersonRei79}, we know that:
\begin{theorem}
The distributed synthesis problem is undecidable.
\end{theorem}

\section{Annotations and Retractions}
Faced with undecidability in general, we ask a different question:
given a game $\calG$ and a specification $\autW$, suppose that
we are told that there exists a winning strategy structure. 
What can we infer from this ? In particular, can we transform the
strategy into one with fewer states, hopefully finitely many ? 

Let us fix a game~$\calG$ described by a family of Mealy machines $M^i$,
and a specification automaton $\autW$. 
An \emph{annotated} strategy for $\calG$ is a structure $(S, \rho)$ 
that expands a strategy structure $S$ 
with a function $\rho: V \to Q^0 \times \dots \times Q^{n-1} \times Q$ labelling every
 node~$v \in V$ with a a profile $(\rho^i ( v ))_{i \in N}$ 
of states of the Mealy machines $M^i$ and a state  $\rho_\autW( v )$
of the tree automaton $\autW$, as follows:
\begin{itemize}
\item  
for any history $\tau$ in $S$ with trace $v_0 \dots v_\ell$, 
the sequence $\rho^i( v_0 ) \dots \rho^i( v_\ell)$ describes the run of $M^i$ on $\tau$;
\item 
if we consider for each history $\tau \in H_S$  
the node~$v_\tau$ reached by the path $v_\init \xrarrow{\tau} v_\tau$ in~$S$,  then  
the mapping $(\rho_\autW( v_\tau ))_{\tau \in H_S}$ describes a run of $\autW$ on $H_S$.
\end{itemize}
A \emph{witness} strategy is an annotated strategy 
where the run described by $\rho_\autW$ is accepting.

Notice that every strategy structure on a tree  
can be expanded with runs of the automata $M^i, \autW$ to obtain an annotated strategy. 
However, for an arbitrary strategy structure~$S$ it may be impossible to 
describe a run as a single-state annotation of its nodes, 
because computations along different paths can reach the same node of~$S$ 
in different states. 
Annotations are indeed runs of vertex-marking automata on (strategy) graphs, 
as studied in \cite{BerwangerJanin06}.
Nevertheless, Rabin's basis theorem~\cite{Rabin69} 
implies that every finite-state winning strategy can be extended -- by taking the synchronised 
product with a certain finite-state automaton -- to allow annotation 
with a witnessing run.

\begin{theorem} Let~$\calG$ be a distributed game with a (finite) set of directions $D$.
\begin{enumerate}[(i)]
\item There exists a winning strategy for $\calG$ if, and only if, there exist a 
witness strategy on the tree $D^*$.    
\item There exists a finite-state winning strategy for $\calG$ if, and only if, 
there exists a witness strategy on a finite set of nodes.
\end{enumerate}
\end{theorem}

\begin{proof}

The if-direction is obvious: 
If a strategy witness $(S, \rho)$ exists, then the underlying 
strategy structure~$S$ can be readily used to solve the
synthesis problem. 

For the converse, suppose that there exists a
winning strategy structure~$S = (V, E, f, v_\init)$ for $\calG$. 
Consider the tree unravelling~$S'$ of $S$, that is, 
the strategy structure on $D^*$ in which the nodes $\pi$ 
correspond to paths in~$S$ 
and the action choice $f'( \pi ) := f( v_\pi )$ is inherited from the end node 
$v_\init \xrarrow{\pi} v_\pi $. 
Since $S$ is a winning strategy, 
the specification automaton $\autW$  has
an accepting run $\rho_\autW: H_S \to Q$ on  
the tree of histories $H_S$ that follow $S$. Further, 
for each player~$i$, the runs of the deterministic Mealy machine $M^i$ on all 
histories that follow $S$ induce a labelling $\rho^i: H_S \to Q^i$. 
For every node $\pi$ in $S'$, set $\rho'( \pi ) = \rho( \tau_\pi )$ 
for the history $\tau_\pi$ identified by the strategy path $\pi$.
Then, $(S', \rho')$ is a witness strategy on $D^*$.
 
In case the witness $S$ at the outset is actually a finite strategy structure, 
the language of witnessing annotations $(S', \rho')$
on the associated strategy tree $S'$ is regular, 
and by Rabin's basis theorem (\cite{Thomas09}),
it contains a regular tree. This can be turned into an annotation of the 
finite-state strategy with the same unravelling.
\end{proof}

Our vehicle for moving from a given strategy structure to one with 
fewer states are particular maps on the strategy nodes.  
For a strategy structure~$S$ and a map $h: V \to V$, 
we define the image $h( S )$ to be the decision structure
$( \hat{V}, \hat{E}, \hat{f}, \hat{v}_\init)$ on a subset $\hat{V} \subseteq V$ 
of nodes with a new transition relation 
$\hat{E} := \{ (u, d, v ) ~|~ (h( u ), d, v) \in E \}$, choice function 
$\hat{f}( v ) := f( h (v ))$, and initial state $\hat{v}_\init = h( v_\init)$; 
we restrict $\hat{V}$ to the set of nodes reachable from $\hat{v}_\init$ via $\hat{E}$.
Intuitively, the map folds a decision structure by redirecting the decision at a node $v$ to the node~$h(v)$ 
and continuing the play from there onwards. This way of performing surgery on strategies is 
used frequently, for instance, 
to show that memoryless strategies are sufficient for winning parity games with perfect information 
(see, e.g.,~\cite{GradelW06}).

A map~$h$ is a \emph{retraction} for an annotated strategy~$(S, \rho)$ 
if\begin{itemize} 
\item $\rho ( v )  = \rho ( h( v ) )$ for all $v \in V$, and
\item $\hat{f}^i( u ) = \hat{f}^i( v )$ 
for every pair of nodes $u \approx^i_{S( h )} v$, for each player~$i$. 
\end{itemize}
The \emph{retract} $h(S, \rho)$ is the 
image $h( S )$ expanded with the annotation $\rho$ 
on the nodes in its domain $\hat{V}$.
 
The constraints on a retraction
to preserve the annotation with states of the observation automata
and to respect the uniformity relation 
ensure that the decision structure obtained as an image is indeed 
a strategy structure.

\begin{lemma} Let $(S, \rho)$ be an annotated strategy for a game~$\calG$. 
If $h$ is a retraction for $(S, \rho)$, then the retract
$h( S, \rho )$ is also an annotated strategy for~$\calG$.
\end{lemma}

We are particularly interested in retractions~$h$ that are \emph{conservative}
in the sense that if $(S, \rho)$ is a witness then the retract $h( S, \rho )$ is also 
a witness.

Safety winning conditions are $\omega$-regular conditions described by 
automata that accept a run only if all the occurring states 
belong to a designated safe subset; 
they can be described by parity automata with only two priorities. 
Since retractions preserve the annotation of strategy nodes 
with states of the specification automaton, 
it immediately follows that retractions are conservative 
for strategies in games with safety conditions. 

\begin{lemma}
Let $\calG$ be a distributed game with a safety winning condition. 
Then, every retraction for a strategy annotation for $\calG$
is conservative.
\end{lemma}

In contrast, liveness properties of strategies
may be hurt by arbitrary retractions. 

\subsection{Progress measures}

We are making some progress in our journey towards identifying the
``essential'' core of the winning strategy, but we are far from done
as yet. The next notion we need is that of {\em progress measures},
as introduced by Klarlund in~his thesis~\cite{Klarlund90}, which yield 
a local representation of an ordering that assigns a 
value to every state of a system, such that by following transitions 
that reduce the value, a specified property is satisfied in the limit.
We use such measures to identify how we might equate states across the temporal
dimension as the strategy evolves over time, such that the parity 
condition is preserved. 

\newcommand{\rank}{r}

For a parity condition $\Omega: Q \to \{0, \dots, r-1 \}$ 
with $\rank$ priorities, 
we consider a standard measure that ranges over $r$-tuples of ordinals. 
Given two tuples   
$x, y \in \On^\rank$ and a priority~$k$,
we write $x \succ_k y$ if $(x_0, \dots, x_k)$ is lexicographically greater than 
$(y_0, \dots, y_k)$.

A \emph{parity progress measure} 
on an annotated strategy $(S, \rho)$ is a 
function $\mu: V \to \On^r$ such that for each transition $(u, d, v)$ in $S$, we have
\begin{itemize}
\item if $k = \Omega( \rho_\autW( u ))$ is even, then $\mu( u ) \succeq_k \mu( v )$, and
\item it $k = \Omega( \rho_\autW( u ))$ is odd, then $\mu( u ) \succ_k \mu( v )$.
\end{itemize}

It is well known that such progress measures 
describe winning strategies in parity games \cite{EJ91,KlarlundK95,Jurdzinski00}.
\begin{theorem}\label{thm:ppm}
An annotated strategy $(S, \rho)$ is a witness if, and only if, 
there exists a parity progress measure on~$(S, \rho)$.
\end{theorem}

\begin{proof}
To see that whenever there exists a parity progress measure for an annotated strategy~$S$, 
the annotation~$\rho_\autW$ describes an accepting run, note that 
each of the preorders~$\succeq_k$ is well-founded (there are no infinite descending chains),
so their lexicographic product $\succeq$ is also well-founded. 
Now suppose that, for some infinite path in~$(S, \rho_\autW)$, 
the least priority~$k$ that occurs infinitely often is odd.  
Since, with every transition $(u, d, w)$ from a node of priority $k = \Omega( \rho_\autW( u ))$ 
in the strategy, the measure $\mu$ decreases in $\succeq_k$ and thus in $\succeq$, this implies that we have an infinite descending chain in $\succeq$ -- a contradiction. 
Accordingly, the run described by $\rho_\autW$ 
satisfies the parity condition.
 
Conversely, let us assume that the given annotation~$(S, \rho)$ describes an accepting run. 
We follow a procedure described by Gr\"adel and Walukiewicz~\cite{GradelW06} 
to define a progress measure $\mu: V \to \On^r$:
For each odd priority~$k$, 
consider the sequence $(Z_\alpha)_{\alpha \in \On}$ of sets, where $Z_0$ consists of
nodes in~$S$ from which 
every path either never reaches priority $k$, or it reaches a smaller priority before reaching~$k$; 
for every ordinal~$\alpha$, 
the set $Z_{\alpha}$ consists of all nodes~$u \in V$ such that, if there is a path from $u$ to a node 
$v$ of priority~$k$, then all successors of $v$ 
belong to $\cup_{\beta < \alpha} Z_{\beta}$. Finally we set~$\mu_k( v )$ to be the 
least ordinal~$\beta$ such that $v \in Z_\beta$. For even priorities~$k$, 
the component~$\mu_k( v )$ is set to zero. 
One can now verify that the mapping $\mu$ defined in this way is a progress measure.
\end{proof}

We say that a retraction $h: V \to V$ is \emph{$\mu$-monotone}, if $\mu( v ) \succeq_k \mu( h( v ))$ 
for every node $v \in V$ of priority $k = \Omega( \rho_\autW( v ))$. 
In addition to preserving runs, monotone retractions preserve progress measures. 
  
\begin{lemma}
Let $(S, \rho)$ be an witness strategy with a parity progress measure~$\mu$. 
Then, any $\mu$-monotone retraction for $(S, \rho)$ is conservative.
\end{lemma}

\begin{proof}
Let~$h$ be a $\mu$-monotone retraction 
for a witness strategy with a parity progress measure as in the statement. 
We show that $\mu$ is a parity progress measure for the retract~$h(S)$ as well:
By definition of the retraction, 
for every transition $(u, d, v)$ in the image $h( S )$, there is a transition 
$(h(u), d, v) \in E$. Moreover, the source nodes of the two transition have the same $\rho$-annotation, 
and hence the same priority~$k$. 
Since $\mu$ is a progress measure on $S$, we have 
$\mu( h(u) ) \succeq_k \mu( v )$, for the priority 
On the other hand, by $\mu$-monotonicity of~$h$, we have $\mu( u ) \succeq_k \mu( h( u ))$. 
Therefore, $\mu(u ) \succeq_k \mu( v )$. In case the priority $k$ of $\rho_\autW(u) = \rho_\autW( h( u)$ is odd, the ordering is strict. Accordingly, $\mu$ is a parity progress measure for the retract $h(S)$ which implies, 
in particular, that $h(S)$ is a witness.
\end{proof}

We can easily verify that $\mu$-monotone retractions are closed under composition. 
\begin{lemma} For a game~$\calG$, suppose there exists a
witnessing strategy $(S, \rho)$ with a parity progress measure $\mu$. 
Let $g$ and $h$ be $\mu$-monotone retractions for $(S, \rho)$ and $h( S, \rho )$, respectively.
Then the composition $g \circ h$ is a $\mu$-monotone retraction for $(S, \rho)$.
\end{lemma}

\section{Compacting Retractions}
Our objective is to retract strategies into smaller ones, hopefully
of finite size.  Towards this, we introduce the notion 
of {\sf distributed states}, which we understand as the atoms of a strategy 
annotation on which retractions will operate.
 
Let us fix an annotated strategy $(S, \rho)$.  Towards defining distributed 
states, it is convenient to include the profile of uniformity relations 
$\approx^i$ into the signature, and to drop the initial state. Thus, we 
shall view the strategy as a structure $S = (V, E, (\equiv^i)_{i \in N}, \rho)$. 
Further, we define the relation $\approx:= \cup_{i \in N}\approx^i$.
 
Now, a distributed state, or shortly \textit{d-state}, is a structure 
$\kappa=(V_\kappa, E,  (\approx^i)_{i \in N}, \rho \}$, 
induced in $S$ by a subset $V_\kappa \subseteq V$ that forms a maximal 
$\approx$-connected component.  Naturally, $E$, $\approx^i$ and $\rho$ 
are the relations of $S$ restricted to $V_\kappa$.  We denote the set of 
all d-states of the structure $S$ by $\mathcal{K}_S$ and the isomorphism 
relation among them by $\cong_{S}$.  Then, $\mathcal{K}_S / \cong_{S}$ is 
the quotient of $\mathcal{K}_S$ by $\cong_{S}$ and $[\kappa]_{\cong_{S}}$ is 
the equivalence class that contains the d-state $\kappa$.

Note that, as defined, it is hardly clear when a d-state is finite.
{\em A priori} the maximal $\approx$-connected components of $S$
would be infinite. This raises the question: when are d-states finite,
and when they are, how does this impact decidability of distributed synthesis.

The notion of a d-state reveals three parameters for the size of a 
witnessing strategy:
the number of connected nodes in single d-state, 
the number of isomorphic d-states in one $\cong$-class, 
and the index of $\cong$, that is, the number of non-isomorphic d-states in $S$. 
For a witnessing strategy to be finite, all these parameters must be finite.

In the rest of the section we show that, 
if all d-states in a witness are of finite size, then, 
for the purposes of decidability, 
the index of $\cong_{S}$, is the only relevant parameter. 
First, we show that for any such witness, we can pick 
an arbitrary $\cong-$class and retract it into a finite one,
without introducing new classes or enlarging the existing ones.

\begin{lemma}\label{lem: finite-U-sets}
Let $S$ be an witness strategy with a parity progress measure $\mu$. Then, for  
any d-state 
$\kappa \in \mathcal{K}_{S}$ of finite size, 
there exists a $\mu$-monotone retraction $h$ 
such that $[\kappa]_{\cong_{h(S)}}$ is a finite set
in the retract $h(S)$ and 
$\mathcal{K}_{h(S)} \subseteq \mathcal{K}_{S}$.
\end{lemma}

\begin{proof}
Let $X := [\kappa]_{\cong_S}$ be the $\cong$-equivalence class of $\kappa$ in $S$. 
For each d-state~$x \in X$, consider an isomorphism $\pi_x: V_\kappa \to V_x$ from $\kappa$, and 
let $\mu_x:V_x \to \On^r$ be the restriction of the parity progress measure $\mu$ to the domain of $x$.
Now consider the point-wise 
ordering on $X$ which puts $x > y$ if $\mu_x \circ \pi_x( v ) \succ \mu_y\circ \pi_y (v )$ 
for each node $v \in V_\kappa$ in the well-founded lexicographic order $\succ$ on $\On^r$.
Since $\kappa$ is finite, by Dickson's Lemma, it follows that $>$ is a well-quasi order.
Now, construct the retraction $h$ that maps every d-state in $X$, 
to the minimum $>$-comparable d-state $X$ 
and fixes the nodes of any d-state that is not in $X$. 
Since $>$ is a well-quasi order, 
the set of incomparable elements is finite.
Therefore $[\kappa]_{\cong_{h(S)}}$ is a finite set. 
As $h$ fixes every $d$-state that is not isomorphic to $\kappa$, we can conclude that 
$h(S)$ contains only $\cong_{h(S)}$-classes of $S$ 
and none of them increase its size.
\end{proof}

If a strategy witness is of finite $\cong$-index and its $d$-states are all finite, 
we can apply the above theorem successively to retract every $\cong$-class into a finite one, 
thus finally obtaining a finite-state strategy. This is the our main result. 

\begin{theorem}[Retraction]\label{thm:finite-finite}
Let $\calG$ be a distributed game with an arbitrary $\omega$-tree regular winning condition. 
If there exists a witness for $\calG$ in which all  d-states are finite and the $\cong$-index is finite as well, 
then there exists a finite winning strategy for~$\calG$. 
\end{theorem}

For a given class~$\calC$ of games, the Retraction Theorem\label{thm:finite-finite} 
can be used as follows. 
We set out by considering tree-shaped strategies for the game instances $\calG \in \calC$. 
Note that in games where the observations and the winning condition
are specified by finite-state automata, 
for every tree-shaped strategy structure~$S$, the d-states $\kappa \in \calK_S$
are finite, since there are only finitely many histories 
of the same length. 
Next, we look at $d$-cells~$\kappa$ that may appear in tree strategies~$S$ 
with a progress measure~$\mu$ and construct partial retractions $h: \kappa \to \kappa$  
that are $\mu$-monotone and image finite;
we speak of \emph{horizontal} retractions, because each~$h$
maps any node of~$S$ to a node of the same depth in the strategy tree.
If we succeed to construct horizontal imaga-finite retractions, we can apply 
Theorem~\ref{thm:finite-finite} -- which, intuitively, states that every $\cong$-class
of a given strategy annotation can be compacted to a finite one via \emph{vertical} 
retractions --
to conclude that 
whenever a game in $\calC$ admits a winning strategy, 
it admits a finite-state winning strategy. 
If, additionally, the construction of horizontal and vertical retractions for a specific 
class~$\calC$
allows to derive 
recursive bounds on the size and number the d-states in the retract, we obtain an effective 
procedure for solving the synthesis problem.   

More generally, the theorem may be used with retractions that are not composed 
of horizontal and vertical mappings, and also with progress measures 
other than parity progress measures.   

\section{Applications}

Theorem~\ref{thm:finite-finite} generalises the perfect-information construction developed
in~\cite{BKP11} for the case of games where the winning condition is \emph{observable}.     
In contrast to our setting where the winning condition automaton~$\autW$
depends on the actions of all players, 
observable winning conditions correspond to the special case where the runs of~$\autW$
depend only on \emph{public} observations. More precisely, $\autW$ corresponds to an automaton that reads 
aggregations $c( b )$ of the observations $b^i$ output by the Mealy automata~$M^i$ such that the 
values $c(b) \neq c(b')$ are different for two profiles~$b$ and~$b'$ 
only if $b^i \neq b'^{i}$ differ for all players~$i$. In other words, for any observation profile~$b$, 
the value $c(b)$ relevant for acceptance is common knowledge among the players.
 
The central result in 
\cite{BKP11} shows that 
there exists a uniform mapping $f: \Gamma^* \to \Gamma^*$ 
on the histories of the game that
induces a retraction for any strategy annotation, 
and moreover guarantees that any two homomorphically equivalent d-states have the same image 
under~$f$. For games with linear-time winning conditions, which can be determinised and thus 
yield a canonical annotation, this leads to retracts in which every homomorphism equivalence type 
of a d-state appears at most once. For games with finitely many d-states, up to homomorphic equivalence, 
this provides an effective solution to the synthesis problem.

Our setting is more general than the one of \cite{BKP11} in two respects: 
winning conditions are formulated as tree properties and, more importantly, 
they may be unobservable. 
Since tree automata recognising the winning condition cannot be determinised, 
there is no canonical run (even in a larger sense, see~\cite{CarayolLNW10}), 
hence our construction relies on fixing a strategy and a run. 
Due to non-observability of the condition, we also need to fix a progress measure -- 
here we opted for parity progress measures, for simplicity; 
other progress measure work as well and may allow a better analysis.
As a consequence of these arbitrary choices, it is not immediate to
obtain general algorithmic results. 
Nevertheless, our framework can be used as a general tool to 
analyse specific game classes. 

For instance, for classes of games in which the size of d-states is 
bounded by the input instance, 
Theorem~\ref{thm:finite-finite} yields an upper bound on 
the size of a minimal winning strategy and thus  
provides a procedure for 
deciding whether a winning strategy exists, and for constructing 
one if this is the case. 
Even if such a procedure would be highly inefficient --
particularly, as it relies on Dickson's lemma -- 
the framework allows to identify decidable instances.

We present two examples to illustrate this approach.
Both refer to slight variations of a standard setting. 
Our first example is on a game between just one player with imperfect information against the 
environment. In the literature, such games have usually been considered
with observable winning conditions (\cite{Reif84,DR11}), 
which in our setting correspond to the situation where the winning condition automaton 
reads the output of the Mealy observation automaton, rather 
than moves that include information about the moves of Nature. 
Moreover, our example refers to branching-time specifications 
instead of the more classical linear-time conditions.
Our second example refers to a multi-player game derived from 
the perfect-information setting by introducing an observation delay.

\subsection{One-player games with hidden objectives}

Let us consider a game~$\calG$ for a single player~$0$ against Nature 
with a winning condition specified by an
$\omega$-tree automaton~$\autW$. 
We will show that whenever there exists a winning strategy for $\calG$, there 
also exists a finite-state winning strategy. 
This is not a new result, it is well known for linear-time winning conditions 
and not surprising in the branching time setting. 

Let us assume that the player 
has a winning strategy, possibly on an infinite set of states. 
Then, there exists 
a tree-shaped witness $\mathcal{S}=(V, E ,v_\init, \rho)$ with a distinct node for every history 
that follows $S$. Further, by Theorem~\ref{thm:ppm}, there exists a 
parity progress measure $\mu$ on $\mathcal{S}$. 
Our aim is to construct a retract $h(S)$
with finitely many d-states, up to isomorphism, and then to apply
Theorem~\ref{thm:finite-finite}.

Note that on any tree-shaped strategy structure, 
the indistinguishability $\sim$ and the uniformity relation~$\cong$ coincide.
Since $S$ is finitely branching, there are only finitely many 
indistinguishable histories of the same length. Hence, 
each~d-state of $S$ is finite. 
As the domain of $S$ is countable, 
we can enumerate the d-states as $\kappa_0, \kappa_1, \dots$

To define the mapping~$h: V \to V$, we consider 
the d-states in this order.  
In the stage corresponding to a d-state $\kappa=\kappa_\ell$, we define  
for every label $q \in Q^1 \times Q_\autW$ that appears on a node of $V_\kappa$ 
the set $U_q := \{u \in V_\kappa~|~\rho( u ) = q \}$ 
and pick the $\succeq$-least element $u_{\min} \in U_q$ with respect to the progress measure~$\mu$.
Then, we map $h( u ) := u_{\min}$ for all nodes $u \in U_q \setminus \cup_{j < \ell} V_{\kappa_j}$ that 
were not previously assigned.

The mapping~$h$ 
defined in this way is a retraction. 
All d-states in the retract are finite -- each label appears at most once, 
hence, their size is at most $(|Q^1| \times |Q|)$. 
Accordingly, there are finitely many~d-states, up to isomorphism. 
By Theorem~\ref{thm:finite-finite}, we can 
thus conclude that every solvable one-player game with imperfect information
admits a finite-state winning strategy.

For the particular case of observable linear-time winning conditions, our construction 
yields the standard powerset construction for solving 
one-player games with imperfect information 
(see, e.g., \cite{DR11}).

\subsection{Coordination games with observation lags}

Our second example involves a team of two players that play 
a parity game against Nature. The setting is standard, we assume that the players 
move in turns and receive perfect information about the current state, 
the only twist is that each of them may receive his information with a delay that is 
nondeterministically chosen by Nature within a bounded time window of~up to $3$ 
rounds (independently for the two players). Such a game model is more
general than that of concurrent games.

Why are such games of interest ? There is a spectrum of game models between
the extremes of perfect information, where every player knows the global
game state, and that of imperfect information, in which a player may
remain perpetually uncertain in his knowledge of global state, or that
of other players. A natural instance of such in-between games would be
one with bounded imperfection, whereby every player receives perfect
information about the global state, up to a bounded delay.

While the detailed formalisation of these games requires some 
redefinitions in our model, it is easy to get the overall
idea of how the techniques developed in this paper can be applied,
and we sketch the idea below.

Consider any witness strategy for the $3$-delay-game, where the underlying
strategy structure is a tree. Clearly such a witness always exists.

We now claim that the tree witness has all d-states of size at most 
$|A|^3$.
To see this, observe that for any two histories $\tau, \tau'$ 
in the witness strategy at depth $\ell$, if their least common ancestor
is at a depth less than $\ell-3$, then they are distinguishable. 
Therefore, every $\approx$-connected component at depth $\ell$ is made 
up of histories which have their least common ancestor at depth greater than or equal 
to $\ell-3$. Since every node has branching factor at most $|\Gamma|$, the 
d-states have size at most $|\Gamma|^3$.

Thus, we have only finitely many non-isomorphic d-states in the
witness tree and hence by our earlier theorem, there is a finite-state
winning strategy as well. Note that this holds for any bounded delay
in receiving perfect information in coordination games.

\section{Discussion}
We have suggested that distributed synthesis in the context of
finite state synchronous coordination games can be studied via 
retractions on winning strategies. The central idea is that
retractions yield finite-state winning strategies when d-states
are themselves finite, and the induced congruence classes can
be bounded. While decision procedures in general work with some
form of quotienting (as in the case of {\em filtrations} employed
in modal logic), imperfect information games bring in the extra
dimension of d-states potentially growing unboundedly. What
is offered here is a technique for combining the two.

That retractions can be composed is easy to see, and hence we can
hope to build structure in strategies via retractions, starting
from abstract ones that realise a limited objective, and refining
them successively. Progress measures then would need to be finer
as well, as the applications demand. The admittedly limited
examples presented here already suggest that there are many
applications ahead.

The main question is a structural characterisation of the ``largest'' 
class of games for which retractions yield finite state winning
strategies, and decidability of the existence of winning strategies.
Another natural question is the characterisation of when memoryless
winning strategies exist. The classification of decidable cases
driven by practical applications (from the viewpoint of system
design and verification) is perhaps more urgent.

\bibliography{global}

\begin{thebibliography}{10}

\bibitem{BerwangerJanin06}
Dietmar Berwanger and David Janin.
\newblock Automata on directed graphs: Vertex versus edge marking.
\newblock In {\em {G}raph {T}ransformations ({ICGT}'06)}, volume 4178 of {\em
  LNCS}, pages 46--60, Natal, Rio Grande do Norte, Brazil, September 2006.
  Springer.

\bibitem{BerwangerKai10}
Dietmar Berwanger and {\L}ukasz Kaiser.
\newblock Information tracking in games on graphs.
\newblock {\em Journal of Logic, Language and Information}, 19(4):395--412,
  October 2010.

\bibitem{BKP11}
Dietmar Berwanger, {\L}ukasz Kaiser, and Bernd Puchala.
\newblock A perfect-information construction for coordination in games.
\newblock In {\em Proceedings of {F}oundations of {S}oftware {T}echnology and
  {T}heoretical {C}omputer {S}cience ({FSTTCS} 2011)}, volume~13 of {\em
  LIPIcs}, pages 387--398. Leibniz-Zentrum f{\"u}r Informatik, December 2011.

\bibitem{BerwangerM14a}
Dietmar Berwanger and Anup~Basil Mathew.
\newblock Infinite games with finite knowledge gaps.
\newblock {\em CoRR}, abs/1411.5820, 2014.
\newblock To appear in Information and Computation.

\bibitem{BerwangerMV15}
Dietmar Berwanger, Anup~Basil Mathew, and Marie Van{ }den{ }Bogaard.
\newblock Hierarchical information patterns and distributed strategy synthesis.
\newblock In {\em {A}utomated {T}echnology for {V}erification and {A}nalysis
  ({ATVA}'15)}, volume 9364 of {\em LNCS}, pages 378--393, Shanghai, China,
  2015. Springer.

\bibitem{CarayolLNW10}
Arnaud Carayol, Christof Loeding, Damian Niwinski, and Igor Walukiewicz.
\newblock {Choice functions and well-orderings over the infinite binary tree}.
\newblock {\em {Central European Journal of Mathematics}}, 8(6):662--682, 2010.

\bibitem{DR11}
L.~Doyen and J.-F. Raskin.
\newblock Games with imperfect information: Theory and algorithms.
\newblock In Krzysztof Apt and Erich Gr{\"a}del, editors, {\em Lectures in Game
  Theory for Computer Scientists}, pages 185--212. Cambridge University Press,
  2011.

\bibitem{EJ91}
E.~Allen. Emerson and Charanjit~S. Jutla.
\newblock {T}ree automata, mu-calculus and determinacy ({E}xtended abstract).
\newblock In {\em 32nd Annual Symposium on Foundations of Computer Science},
  pages 368--377, San Juan, Puerto Rico, 1--4 October 1991. IEEE.

\bibitem{FinkbeinerSch05}
B.~Finkbeiner and S.~Schewe.
\newblock {U}niform distributed synthesis.
\newblock In {\em Proc. of Logic in Computer Science~(LICS'05)}, pages
  321--330. IEEE, 2005.

\bibitem{GastinLerZei04}
Paul Gastin, Benjamin Lerman, and Marc Zeitoun.
\newblock {D}istributed games and distributed control for asynchronous systems.
\newblock In {\em Proc. Latin American Theoretical Informatics Symposium
  (LATIN'04)}, number 2976 in LNCS, pages 455--465. Springer, 2004.

\bibitem{GradelW06}
Erich Gr{\"{a}}del and Igor Walukiewicz.
\newblock Positional determinacy of games with infinitely many priorities.
\newblock {\em Logical Methods in Computer Science}, 2(4), 2006.

\bibitem{Henzinger05}
Thomas~A. Henzinger.
\newblock {G}ames in system design and verification.
\newblock In {\em Proc. Theoretical Aspects of Rationality and Knowledge
  ({TARK}-2005)}, 2005.

\bibitem{Janin07}
David Janin.
\newblock On the (high) undecidability of distributed synthesis problems.
\newblock In {\em Proc. of Theory and Practice of Computer Science (SOFSEM
  2007)}, volume 4362 of {\em LNCS}, pages 320--329. Springer, 2007.

\bibitem{Jurdzinski00}
M.~Jurdzi{\'n}ski.
\newblock {S}mall {P}rogress {M}easures for {S}olving {P}arity {G}ames.
\newblock In {\em Symposium on Theoretical Aspects of Computer Science (STACS
  2000), Proceedings}, volume 1770 of {\em LNCS}, pages 290--301. Springer,
  2000.

\bibitem{Klarlund90}
N.~Klarlund.
\newblock {\em Progress Measures and Finite Arguments for Infinite
  Computations}.
\newblock Number no. 1153 in Progress measures and finite arguments for
  infinite computations. Cornell University, Department of Computer Science,
  1990.

\bibitem{Klarlund94}
Nils Klarlund.
\newblock Progress measures, immediate determinacy, and a subset construction
  for tree automata.
\newblock {\em Annals of Pure and Applied Logic}, 69(2):243--268, 1994.

\bibitem{KlarlundK95}
Nils Klarlund and Dexter Kozen.
\newblock Rabin measures.
\newblock {\em Chicago J. Theor. Comput. Sci.}, 1995, 1995.

\bibitem{KupfermanVar01}
Orna Kupferman and Moshe~Y. Vardi.
\newblock Synthesizing distributed systems.
\newblock In {\em Proc. of LICS~'01}, pages 389--398. IEEE Computer Society
  Press, June 2001.

\bibitem{MadhuThiagu02}
P.~Madhusudan and P.S. Thiagarajan.
\newblock A decidable class of asynchronous distributed controllers.
\newblock In {\em Concurrency Theory (CONCUR 2002) Proceedings}, volume 2421 of
  {\em Lecture Notes in Computer Science}, pages 145--160. Springer Berlin
  Heidelberg, 2002.

\bibitem{McNaughton66}
R.~McNaughton.
\newblock {T}esting and generating infinite sequences by a finite automaton.
\newblock {\em Information and Computation}, 9:521--530, 1966.

\bibitem{MohalikWal03}
Swarup Mohalik and Igor Walukiewicz.
\newblock {D}istributed {G}ames.
\newblock In {\em FSTTCS'03}, volume 2914 of {\em LNCS}, pages 338--351, 2003.

\bibitem{MuschollWal14}
Anca Muscholl and Igor Walukiewicz.
\newblock Distributed synthesis for acyclic architectures.
\newblock In {\em Foundation of Software Technology and Theoretical Computer
  Science, {FSTTCS} 2014, Proc.}, volume~29 of {\em LIPIcs}, pages 639--651.
  Schloss Dagstuhl - Leibniz-Zentrum fuer Informatik, 2014.

\bibitem{PetersonRei79}
Gary~L. Peterson and John~H. Reif.
\newblock {M}ultiple-{P}erson {A}lternation.
\newblock In {\em Proc 20th Annual Symposium on Foundations of Computer
  Science, (FOCS 1979)}, pages 348--363. IEEE, 1979.

\bibitem{PnueliRos89}
A.~Pnueli and E.~Rosner.
\newblock {O}n the synthesis of a reactive module.
\newblock In {\em Proceedings of the 16th ACM SIGPLAN-SIGACT symposium on
  Principles of programming languages}, pages 179 -- 190. ACM Press, 1989.

\bibitem{PnueliRos90}
Amir Pnueli and Roni Rosner.
\newblock Distributed reactive systems are hard to synthesize.
\newblock In {\em Proceedings of the 31st Annual Symposium on Foundations of
  Computer Science, FoCS~'90}, pages 746--757. IEEE, 1990.

\bibitem{Rabin69}
M.~Rabin.
\newblock {D}ecidability of second-order theories and automata on infinite
  trees.
\newblock {\em Transactions of the AMS}, 141:1--35, 1969.

\bibitem{Rabin72}
Michael~Oser Rabin.
\newblock {\em Automata on Infinite Objects and Church's Problem}.
\newblock American Mathematical Society, Boston, MA, USA, 1972.

\bibitem{RamSim10}
R.~Ramanujam and Sunil~Easaw Simon.
\newblock A communication based model for games of imperfect information.
\newblock In {\em Proc. of CONCUR~'10}, volume 6269 of {\em LNCS}, pages
  509--523. Springer, 2010.

\bibitem{Reif84}
J.~Reif.
\newblock {T}he complexity of two-player games of incomplete information.
\newblock {\em Journal of Computer and System Sciences}, 29:274--301, 1984.

\bibitem{Thomas09}
Wolfgang Thomas.
\newblock Facets of synthesis: Revisiting {C}hurch's {P}roblem.
\newblock In {\em Proceedings of the 12th International Conference on
  Foundations of Software Science and Computational Structures (FOSSACS '09)},
  pages 1--14. Springer, 2009.

\bibitem{VoegeJ00}
Jens V{\"{o}}ge and Marcin Jurdzi{\'{n}}ski.
\newblock A discrete strategy improvement algorithm for solving parity games
  ({E}xtended abstract).
\newblock In {\em Computer Aided Verification, CAV 2000, Proceedings}, volume
  1855 of {\em LNCS}, pages 202--215, Chicago, IL, USA, July 2000.
  Springer-Verlag.

\end{thebibliography}

\end{document}